# Aharonov-Bohm Oscillations in a Quasi-Ballistic 3D Topological-Insulator Nanowire

S. Cho[†1], B. Dellabetta[2], R.D. Zhong[3], J. Schneeloch[3], T.S. Liu[3], G. Gu[3], M.J. Gilbert[2], N. Mason[*1]

**In three-dimensional topological insulators (3D TI) nanowires, transport occurs via gapless surface states where the spin is fixed perpendicular to the momentum[1-6]. Carriers encircling the surface thus acquire a $\pi$ Berry phase, which is predicted to open up a gap in the lowest-energy 1D surface subband. Inserting a magnetic flux ($\Phi$) of $h/2e$ through the nanowire should cancel the Berry phase and restore the gapless 1D mode[7-8]. However, this signature has been missing in transport experiments reported to date.[9-11] Here, we report measurements of mechanically-exfoliated 3D TI nanowires which exhibit Aharonov-Bohm oscillations consistent with topological surface transport. The use of low-doped, quasi-ballistic devices allows us to observe a minimum conductance at $\Phi = 0$ and a maximum conductance reaching $e^2/h$ at $\Phi = h/2e$ near the lowest subband (i.e. the Dirac point), as well as the carrier density dependence of the transport.**

The band structure of surface carriers in a 3D TI nanowire is described by a 1D momentum vector $k$ along the nanowire axis. When a magnetic field ($B$) is applied along the nanowire axis, the surface electron picks up an additional phase of $2\pi\Phi/\Phi_0$, where $\Phi_0 = h/e$ is the magnetic flux quantum. Accordingly, the 1D subbands are described by:

[1] Department of Physics and Frederick Seitz Materials Research Laboratory, 104 South Goodwin Avenue, University of Illinois, Urbana, Illinois 61801, USA. email: †sjcho11@illinois.edu, * nadya@illinois.edu.
†current address: Department of Physics, Korea Advanced Institute of Science and Technology, Daejeon 305-701, Korea
[2] Department of Electrical and Computer Engineering and Micro and Nanotechnology Laboratory, University of Illinois, 208 N. Wright St., Urbana, IL 61801, USA.
[3] Condensed Matter Physics and Materials Science Department, Brookhaven National Laboratory, Upton, NY 11973, USA

$$E_{kl} = \pm \hbar v_F \sqrt{k^2 + \frac{\pi \left(l + \frac{1}{2} - \Phi/\Phi_0\right)^2}{S}}$$

(1)

Here, $l + \frac{1}{2} = \pm \frac{1}{2}, \pm \frac{3}{2}, \pm \frac{5}{2}, ...$ is a (half integer) angular momentum, $v_F$ is the Fermi velocity, and $S$ is the cross-sectional area of the nanowire[10]. The energy-momentum relation is clearly periodic in $\Phi/\Phi_0$, leading to $h/e$ Aharonov-Bohm (AB) oscillations[12-14]. The topological nature of the 1D surface subbands can be observed via the behavior of the oscillation maxima and minima as a function of $\Phi/\Phi_0$. Fig. 1a depicts schematic band structures for $\Phi/\Phi_0 = 0$, 0.1, and 0.5. The 1D subbands are gapped at $\Phi/\Phi_0 = 0$ by $\Delta$. While all the subbands at $\Phi/\Phi_0 = 0$ are doubly-degenerate, the degeneracy becomes lifted at $\Phi/\Phi_0 \neq 0$. A maximum in conductance occurs at $\Phi/\Phi_0 = 0.5$ when the magnetic flux cancels the $\pi$ Berry phase, and a non-degenerate gapless lowest-energy mode appears[7-8]. This mode contains massless Dirac-like excitations that follow a 1D linear energy-momentum dispersion, and is not observed in non-topological systems[12-15].

The behavior of the AB oscillations should also vary with carrier density[16], which can be tuned by an external voltage. In Fig. 1b, a schematic mapping shows the number of modes at various Fermi energies, $E_F$. At $|E_F| < \Delta/2$, the number of modes at $\Phi/\Phi_0 = 0$ and 0.5 should be 0 and 1 respectively. Therefore, in a ballistic transport regime, the conductance should exhibit a minimum at $\Phi/\Phi_0 = 0$ and a maximum $\approx e^2/h$ at $\Phi/\Phi_0 = 0.5$. Away from the Dirac point the phase of the conductance alternates: when $E_F$ is located at one of the green(red)-dotted lines in Fig. 1b, the conductance should have a minimum(maximum) at $\Phi/\Phi_0 = 0$ and a maximum(minimum) at $\Phi/\Phi_0 = 0.5$, as depicted in Fig. 1c.

Previous experiments have demonstrated AB oscillations in 3D TI nanowires, indicating surface transport[9-11] However, the predicted behavior close to the Dirac point has not been observed. The experiments performed here differ in several significant ways. First, excess n-doping was removed using the chemical dopant 2,3,5,6-tetrafluoro-7,7,8,8-tetracyanoquinodimethane (F4-TCNQ), allowing transport to be measured while $E_F$ is tuned through the Dirac point[1,17]. Second, the small sample sizes used here allow for quasi-ballistic transport. In the quasi-ballistic regime, $h/e$ AB oscillations are predicted to dominate over $h/2e$ Aharonov-Altshuler-Spivak (AAS) oscillations[7,14,15]. AAS oscillations originate from interference between time-reversed paths due to weak (anti)localization and thus differ from AB oscillations in their physical origin. In the quasi-ballistic regime, the channel length ($L$) should not be much longer than the mean-free path ($l_m$), which is ~ 300 nm in $Bi_2Se_3$ nanowires. Additionally, $L$ should be long enough for the electrons to circumnavigate the nanowire before exiting the channel, in order to observe dependence of the conductance on a magnetic flux.

Our samples consist of nanowires obtained via mechanical exfoliation of a bulk TI crystal $(Bi_{1.33}Sb_{0.67})Se_3$ onto $Si/SiO_2$ substrates, where the doped substrate acts as a global backgate (see Methods). While chemical or mechanical etching allow effective production of nanowires from layered thin-films, these processes typically introduce additional damage and defects along the edges[18-19]. We fabricated two four-terminal devices having channel lengths of $L_1 = 200$ nm (device 1) and $L_2 = 350$ nm (device 2) (see Fig. 2). Device 1 has width = 110 nm and thickness = 15 nm ($S_1 = 1.65 \times 10^{-15}$ m$^2$), while device 2 has width = 100 nm and thickness = 16 nm ($S_2 = 1.60 \times 10^{-15}$ m$^2$). The typical mean free path $l_m \approx 100$ nm of the nanowires is estimated from the gate-dependent conductivity of the surface electrons using $l_m = v_F \tau$ and $\tau = \hbar \sigma \left(\frac{\pi}{n}\right)^{\frac{1}{2}} / (e^2 v_F)$. Thus, both devices are in the quasi-ballistic regime where $L/l_m$ ~ 2-3 [20]. Samples were measured using standard four-terminal lock-in techniques in a dilution refrigerator at the base temperature of 16 mK. Figure 2c shows the location of the Dirac point for device 1 at $V_g \approx$ -15V, and demonstrates that the carrier density can be tuned through the Dirac point with a backgate voltage. It has been shown that when the initial doping level is low in nanostructured TI devices, the Fermi energies of the top and bottom surfaces could be simultaneously tuned with a single gate-electrode, due to the large inter-surface capacitance[1].

The magneto-conductance at the Dirac point of device 1 is plotted as a function of $\Phi/\Phi_0$ in Fig. 3a. The magneto-conductance clearly shows $h/e$ AB oscillations, for flux through an area consistent with the measured cross section of the nanowire. The fast Fourier Transform of the data (Fig. 3b) confirms the predominant $h/e$ AB and absent $h/2e$ AAS oscillations. The magneto-conductance minimum near $\Phi/\Phi_0 = 0$ and maximum near $\Phi/\Phi_0 = 0.5$ are consistent with the existence of a low-energy topological mode. Moreover, maximum conductance values close to $e^2/h$ ($\approx 0.9\ e^2/h$) at $\Phi/\Phi_0 = 0.5$ imply quasi-ballistic transport within this mode[7,8,21]. Quasi-ballistic behavior is further confirmed by the observation of Fabry-Perot oscillations in similar TI nanowire devices[22-24] (see Supplementary Discussion A). Similar magneto-conductance minimum at $\Phi/\Phi_0 = 0$ and maximum $\approx e^2/h$ at $\Phi/\Phi_0 = 0.5$ were observed in device 2, as well as for device 1 in different thermal cycles (see Supplementary Discussion B). This reproducibility suggests that the behavior of the AB oscillations at the Dirac point is a robust phenomenon.

Figure 3c shows the magneto-conductance for increasing gate voltages, demonstrating that the AB phase alternates with an approximate period of $\Delta V_g = 2$V. This is consistent with the phase alternations predicted to occur periodically when $E_F$ increases by ½ $\Delta$[7-8]. Similar phase alternations with gate voltage were observed in device 2 (see Supplementary Discussion B). Due to the limited resolution in gate voltage steps (limited by voltage jumps due to local charging effects), Fig. 3c may not show all the alternating phases of the AB oscillations; however, the overall dependence of the magneto-conductance on gate voltage is qualitatively consistent with theoretical expectations[7-8]. Because subbands away from the Dirac point may be diffusive, the amplitudes of AB oscillations in these regions vary between 0 and $e^2/h$ [20].

The conductance maxima near the Dirac point deviate slightly from $e^2/h$, which can be explained by the fact that, when current is flowing in a 3D TI nanowire (and thus time-reversal symmetry is broken by the non-equilibrium state), the conductance is not guaranteed to be quantized[2,4]. The deviation may also be partly due to the inevitable invasiveness of voltage probes in 1D systems[20]. The data also show finite conductance ($\approx 0.5\ e^2/h$ in Fig. 3a) at $\Phi/\Phi_0 = 0$ in the gapped nanowire near the Dirac point. This could occur if chemical potential fluctuations due to strong disorder were much larger than the gap size[7-8]. However, in that case, $h/2e$ AAS oscillations would dominate over $h/e$ AB oscillations, which is inconsistent with our observations. Rather, the finite conductance at $\Phi/\Phi_0 = 0$ is likely due to small deviations of $E_F$ from the Dirac point, as discussed below. The data further show that the phase of the AB oscillations becomes less regular at large magnetic field, which is likely due to the Zeeman energy. At low magnetic field, the Zeeman energy is negligible compared to the energy spacing of subbands: $\Delta E_{Zeeman} = \frac{ge\hbar\Phi_0}{4m_e S_1} \approx 1.66$ meV $\ll \Delta_1 = \hbar v_F \sqrt{\pi/S_1} \approx 14.4$ meV in device 1 at $\Phi/\Phi_0 = 0.5$, where g $\approx 23$ is the g-factor. However, at large $B$ ($\Phi/\Phi_0 > 2$) the Zeeman energy is comparable to $\Delta$.

The magneto-conductance data can be better understood by comparing it to full 3D simulations of TI nanowires coupled to metallic contacts, calculated within the non-equilibrium Green's function formalism[24] (see Supplementary Discussion C). Figure 4 plots simulated magneto-conductance for a series of $E_F$ ranging from the Dirac point to $\Delta$, demonstrating the AB phase alternation with increasing $E_F$. Figure 4a shows finite zero-field conductance due to off-resonant transmission, which can occur when short wires are coupled to metallic contacts. The simulations indicate that increasing $E_F$ only a few meV from the Dirac point can add a significant fraction of $e^2/h$ to zero-field conductance. For example, $G(\Phi/\Phi_0 = 0)$ increases from 0.20 $e^2/h$ to 0.45 $e^2/h$ when $E_F$ shifts from the Dirac point to $\Delta/4$ (a shift of ≈ 3.6 meV for device 1). In addition, zero-field conductance can be lifted by any band bending brought about by the chemical dopant, which can effectively add a linear chemical potential across the thickness of the nanowire. Though the chemical doping method we employ is not expected to produce significant band bending in $Bi_2Se_3$[1], the nanowire magneto-conductance is very sensitive to slight deviations in chemical potential, due to the small surface state band gap. For example, Fig. 4b shows that band bending of 30 meV ≈ $2\Delta$ can yield finite conductance at $\Phi/\Phi_0 = 0$, at both the Dirac point and $\Delta/4$. Although it is difficult to distinguish the relative experimental magnitude of each contribution, the Figure shows that a combination of these two effects can fully account for the zero-field conductance near the Dirac point, and leads to a good qualitative agreement between theory and experiment.

    By fabricating quasi-ballistic 3D TI nanowire devices that are gate-tunable through the Dirac point, we have been able to demonstrate salient features of AB oscillations not seen in other non-topological nanowire systems. In particular, we observe alternations of conductance maxima and minima with gate voltage, and conductance minima near $\Phi/\Phi_0 = 0$ with corresponding maxima of ~ $e^2/h$ near $\Phi/\Phi_0 = 0.5$, which is consistent with the existence of a low-energy topological mode. The observation of this mode is a necessary step toward utilizing topological properties at the nanoscale in post-CMOS applications, for example, in topological quantum computing devices[25] or as efficient replacements for metallic interconnects in information processing architectures[26].

# Methods

Three-dimensional TI crystals ($Bi_{1.33}Sb_{0.67}Se_3$) were grown using a modified floating zone method[16]. TI nanowires were obtained by mechanical exfoliation ("scotch tape method") of bulk ($Bi_{1.33}Sb_{0.67}$)$Se_3$ crystals on 300nm $SiO_2$/highly *n*-doped Si substrates[27]. The nanowires were identified via optical and atomic force microscopy (AFM)[1,28,29]. Nanowire thickness was determined by AFM, while the width was more accurately determined by scanning probe microscopy after all electrical measurements were completed. After the nanowires were located on the $SiO_2$/Si chips, electron beam lithography was performed to define four-point electrodes. Subsequently, the surface was cleaned with brief ion milling and Ti(2.5nm)/Au(50nm) were deposited at a base pressure $\approx 1 \times 10^{-9}$ Torr. Immediately after lift-off, ~ 14 nm of F4-TCNQ (Sigma-Aldrich) was deposited via thermal evaporation. Finally, the devices were wire-bonded and cooled down in a commercial dilution refrigerator. All the electrical measurements were performed at a base temperature $\approx$ 16 mK by using standard ac lock-in techniques.

**Acknowledgements**

NM, SC, BD and MJG acknowledge support from the ONR under grant N0014-11-1-0728. MJG acknowledges support from the ONR under grant N0014-11-1-0123 and NSF under grant CAREER ECCS-1351871. Device fabrication was carried out in the MRL Central Facilities (partially supported by the DOE under DE-FG02-07ER46453 and DE-FG02-07ER46471). The work at BNL was supported by the US Department of Energy, Office of Basic Energy Sciences, under contract DE-AC02-98CH10886. SC acknowledges useful discussions with J. Bardarson and P. Kim.

**Author contributions**

SC fabricated the devices and performed the electrical measurements. AY, JS, ZJX, and GG grew the bulk TI crystal. SC, BD, MJG and NM analyzed the data and wrote the paper together.

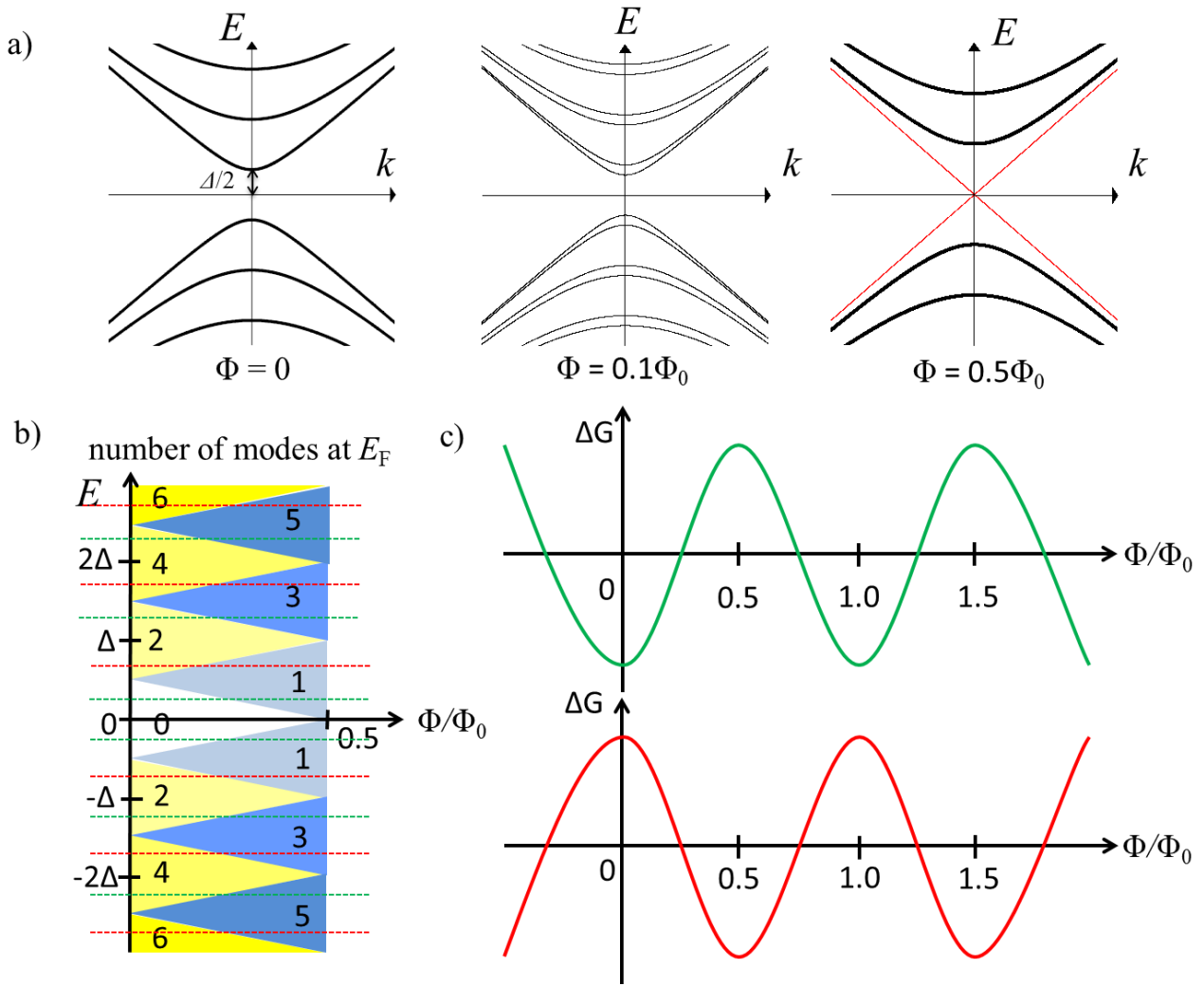

**Figure 1| Schematics of TI nanowire band structure and magneto-conductance.** (a) Schematic band structure of a TI (Bi$_2$Se$_3$) nanowire at $\Phi/\Phi_0 = 0, 0.1, 0.5$, showing the appearance of the low-energy topological mode at $\Phi/\Phi_0 = 0.5$. The subband spacing, $\Delta$, for $\Phi = 0$ and $k = 0$ is labeled. (b) 1D subband occupancy for varying chemical potential, $E_F$, as a function of $\Phi/\Phi_0$. The occupancy increases with $\Delta$. (c) Expected alternating patterns of magneto-conductance oscillations with $E_F$. Green and red lines curves correspond to same color dashed lines in (b).

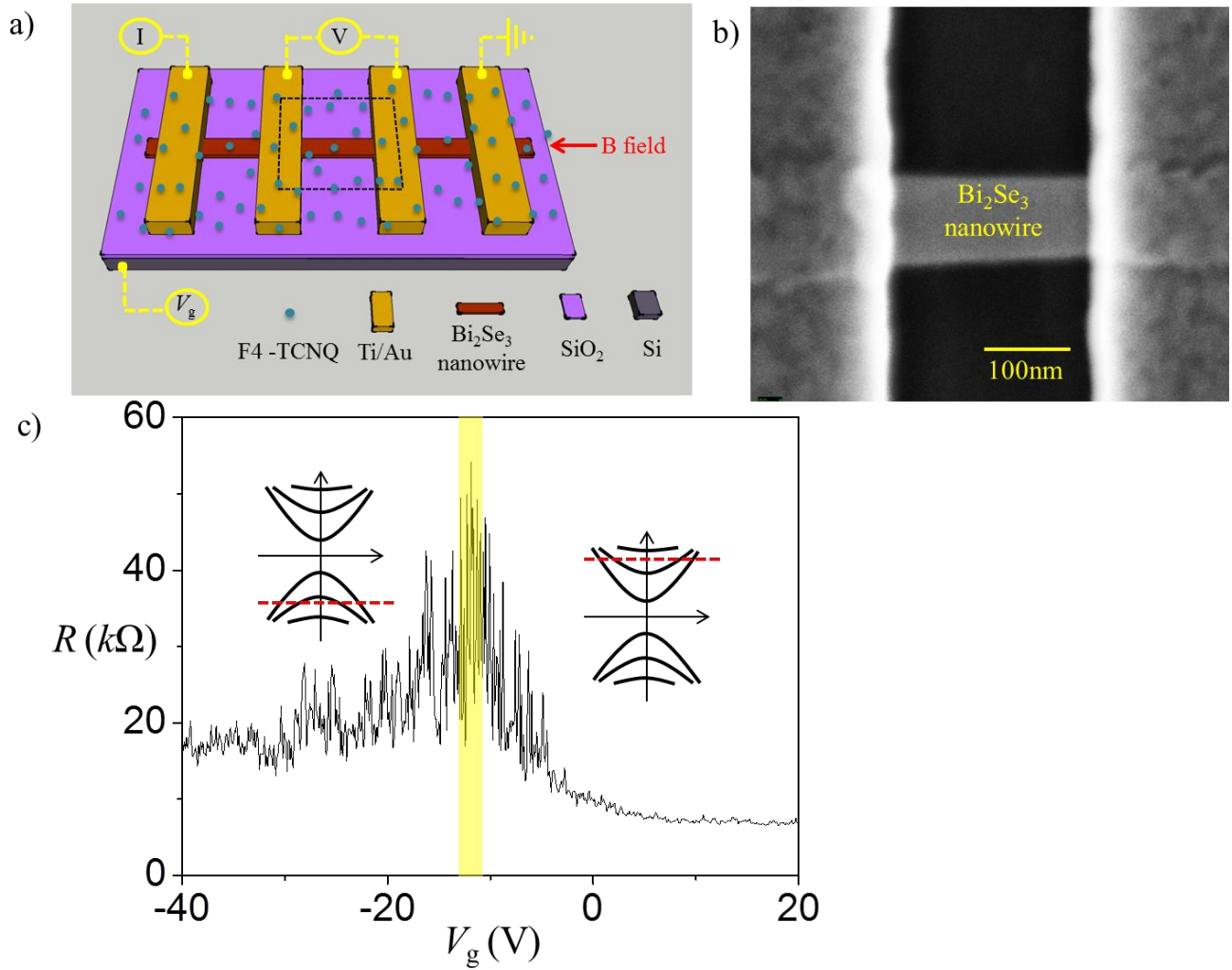

**Figure 2| Device configuration and characterization. (a)** Device and measurement schematic showing four-point lead configuration and chemical dopant, F4-TCNQ. **(b)** Scanning Electron Microscope image of device 1, corresponding to dashed region of (a) and **(c)** Four-terminal resistance $R$ as a function of back-gate voltage $V_g$ at $B = 0$ and $T = 16$ mK. The resistance maximum indicates the location of the Dirac point at $V_g \approx -12$ V (yellow shaded region: $-13$V $< V_g < -11$V). Inset schematics indicate the position of $E_F$ in either the conduction or valence band.

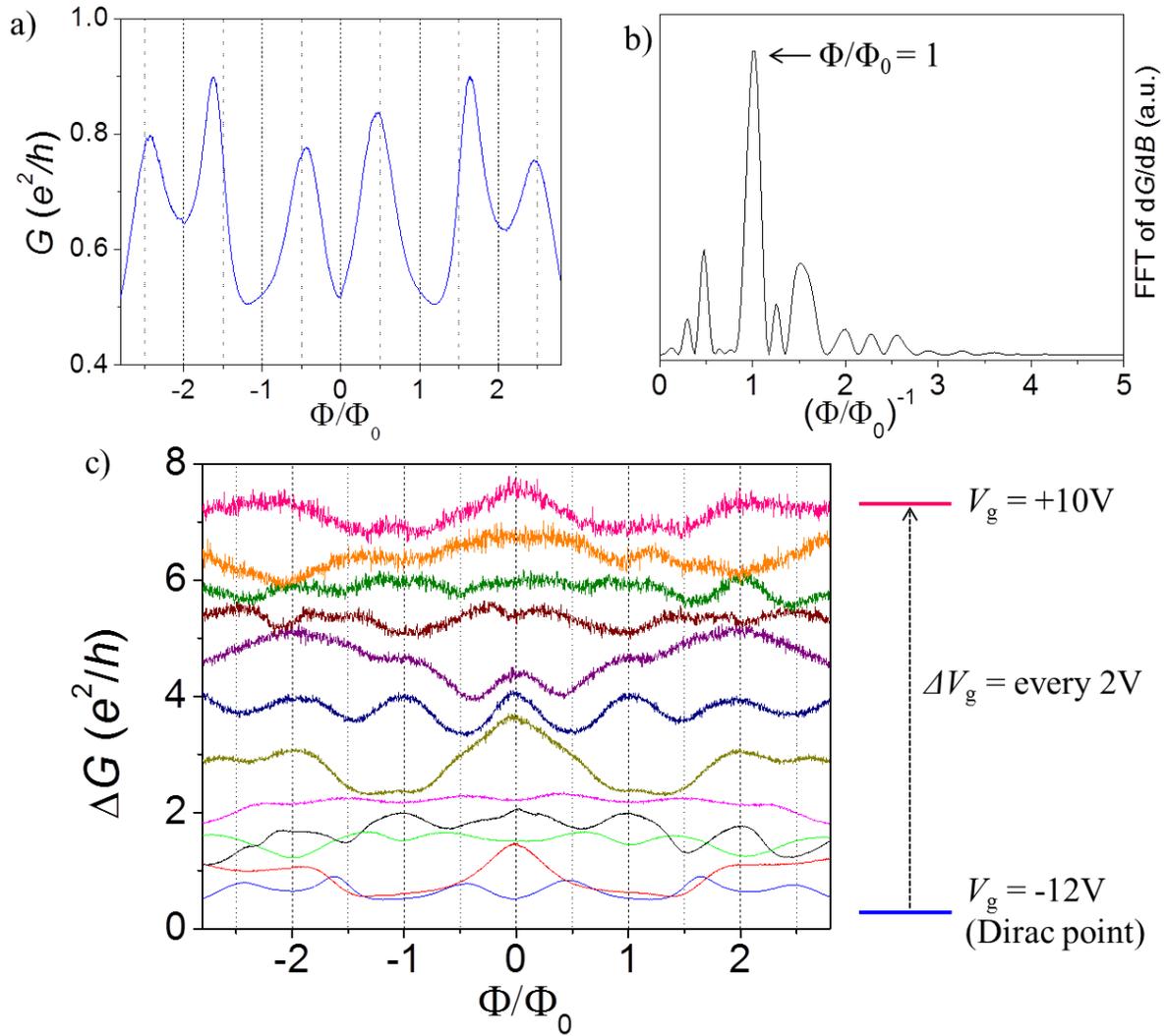

**Figure 3| Magneto-conductance data for device 1.** (**a**) Magneto-conductance measured at the Dirac point ($V_g \approx -12$ V), showing AB oscillations with a period of $\Phi/\Phi_0 = 1$. A conductance minimum occurs at $\Phi/\Phi_0 = 0$. (**b**) Fast Fourier Transform of $dG/d(\Phi/\Phi_0)$ from (a) showing dominant AB ($h/e$) and lack of AAS ($h/2e$) oscillations. (**c**) Magneto-conductance measured away from the Dirac point showing alternating phase of AB with gate voltage.

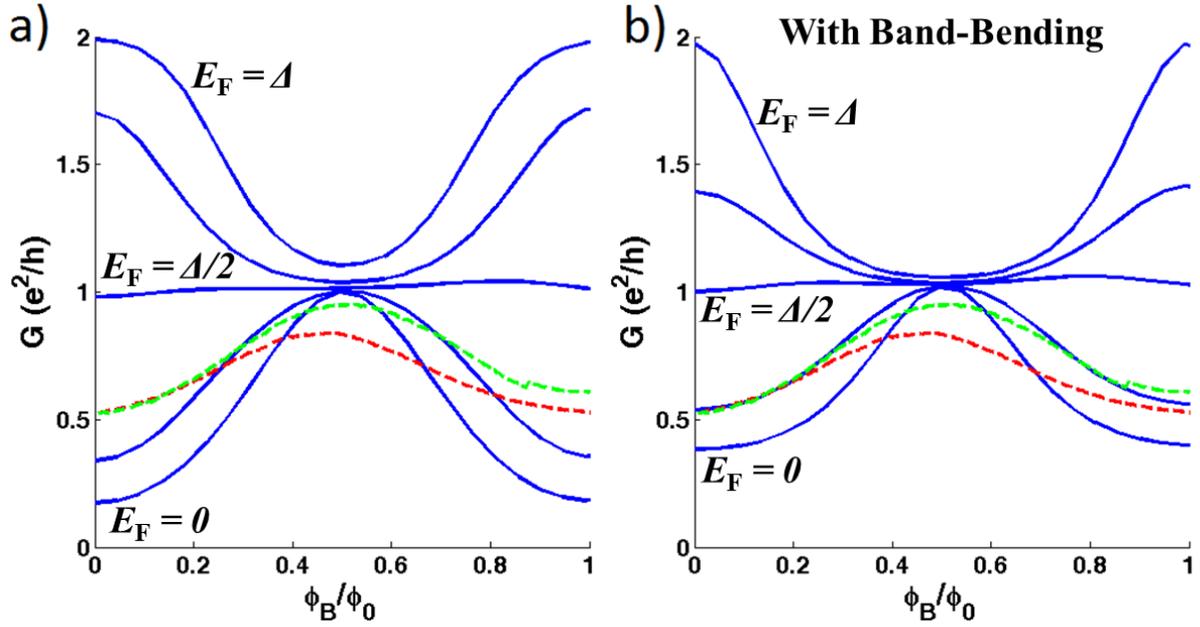

**Figure 4| Simulated magneto-conductance. (a)** AB oscillations for a TI nanowire for $E_F$ ranging from the Dirac point to $\Delta$ in increments of $\Delta/4$. The conductance at $G(\Phi/\Phi_0 = 0)$ increases with increasing $E_F$. The Dirac point conductance $G(\Phi/\Phi_0 = 0) = 0.20\ e^2/h$ emerges due to evanescing modes from the metallic contacts, though this effect is enhanced in the simulations due to the short channel length (see Supplementary Discussion C). Experimental magneto-conductance from device 1 (red dashed) and device 2 (green dashed) are overlaid. **(b)** AB oscillations for a TI nanowire with a linear band bending profile of $2\Delta$ for the same $E_F$ values as (a) showing a further increase in conductance at $G(\Phi/\Phi_0 = 0)$. The simulations show good qualitative agreement with the overlaid experimental data.

**Supplementary Discussion**

Supplementary Discussion A: Fabry-Pérot interference in a 3D TI nanowire device

Figure S1 shows 2D differential conductance plotted as a function of source-drain voltage $V_{sd}$ and gate voltage $V_g$ for a 3D TI nanowire device having dimensions similar to that of device 1 (channel length $L \approx 200$nm). The 2D conductance plot clearly shows resonant patterns of diagonal lines, as well as alternating conductance maxima and minima at $V_{sd}=0$, all at high conductance, which are signatures of Fabry–Pérot interference. Here, the TI nanowire device acts as a Fabry-Pérot cavity for electrons confined between partially-transmitting metal leads. The observation of Fabry-Pérot interference indicates that ballistic electron transport occurs through resonant states in the TI nanowire cavity. Similar interference patterns have been reported in graphene[1-2] and carbon nanotubes[3]. Note that the possibility of Coulomb blockade can be excluded since the conductance is much greater than $e^2/h$ at low temperature, and the diamond patterns associated with suppressed conductance at $V_{sd}=0$ are missing. The observation of Fabry-Pérot oscillations in a TI nanowire nearly identical to device 1 supports the interpretation that quasi-ballistic electron transport occurs in device 1.

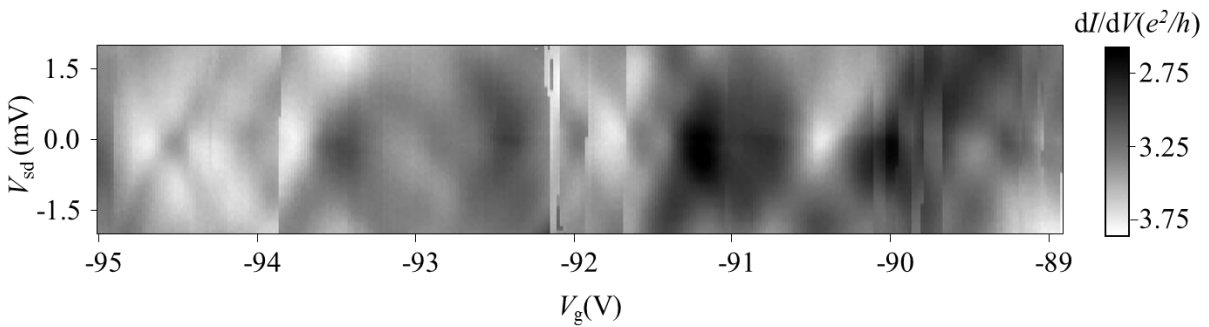

**Figure S1| Fabry-Perot Oscillations.** 2D plot of differential conductance d$I$/d$V$ as a function of bias voltage $V_{sd}$ and gate voltage $V_g$ measured in a TI nanowire device having dimensions (length $\approx 200$ nm, width $\approx 90$ nm and thickness $\approx 13$ nm ) comparable to device 1. Sharp vertical lines are due to switching events related to substrate charging effects.

Supplementary Discussion B: Reproducibility and length-dependent AB oscillations

Here we discuss our additional data obtained in device 1 ($L_1$ = 200 nm) before thermal cycling and in device 2, which has longer channel length ($L_2$ = 350 nm). The dimensions of the nanowires in the devices are comparable: device 1 has width = 110 nm and thickness = 15 nm ($S_1 = 1.65 \times 10^{-15}$ m$^2$), while device 2 has width = 100 nm and thickness = 16 nm ($S_2 = 1.60 \times 10^{-15}$ m$^2$). In Fig. S2a, the magneto-conductance at the Dirac point of device 1, obtained before thermal cycling, is plotted as a function of $\Phi/\Phi_0$. In general, thermal cycling should scramble any random aspects of the magneto-conductance phase and behavior. The magneto-conductance and the fast Fourier Transform (FFT) in Fig. S2b clearly show $h/e$ AB oscillations, and are similar to what was obtained after thermal cycling (see the main article). In particular, the magneto-conductance exhibits a minimum at $\Phi/\Phi_0 = 0$ near the Dirac point and a maximum at $\Phi/\Phi_0 = 0.5$, as well as an alternating AB phase as a function of gate voltage (Fig. S2d). Figure S3 shows that similar behavior occurs in device 2—i.e., a magneto-conductance minimum at $\Phi/\Phi_0 = 0$ and a maximum $\approx e^2/h$ at $\Phi/\Phi_0 = 0.5$ observed at the Dirac point ($V_g$ = -31V), along with alternating AB phase as a function of gate voltage. This reproducibility suggests that the behavior of the AB oscillations at the Dirac point is a robust phenomenon. Figure S3b shows that both $h/e$ AB and $h/2e$ AAS oscillations (1 < $\Phi/\Phi_0$ < 2) occur in device 2. Considering that device 2 has a longer channel length ($L_2$ = 350nm) than device 1 ($L_1$ = 200nm), the observation of both AAS and AB oscillations imply that device 2 is at the border between a quasi-ballistic and a diffusive regime[4,5]. However, the fact that few AAS oscillations appear in magneto-conductance oscillations away from the Dirac point (Fig. S3d) may indicate that the transport regime is closer to a quasi-ballistic regime than to a diffusive one.

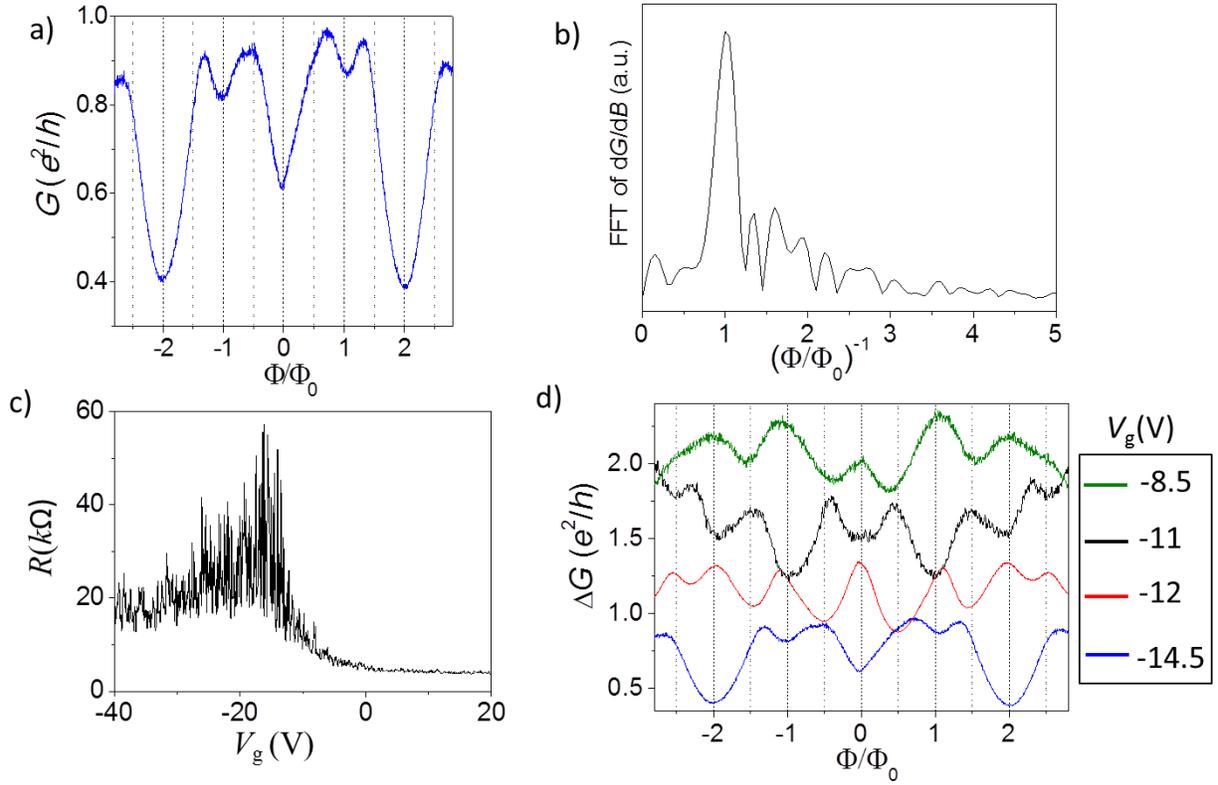

**Fig. S2| AB oscillations in device 1 before thermal cycling.** (**a**) Magneto-conductance at the Dirac point (Vg = -14.5V), showing a conductance minimum at $\Phi/\Phi_0 = 0$. (**b**) FFT of dG/dB from the data in (a), showing dominant AB ($h/e$) oscillations. (**c**) Gate dependent four-point resistance, showing Dirac point. (**d**) Magneto-conductance measured for increasing gate voltages, showing alternating AB phase.

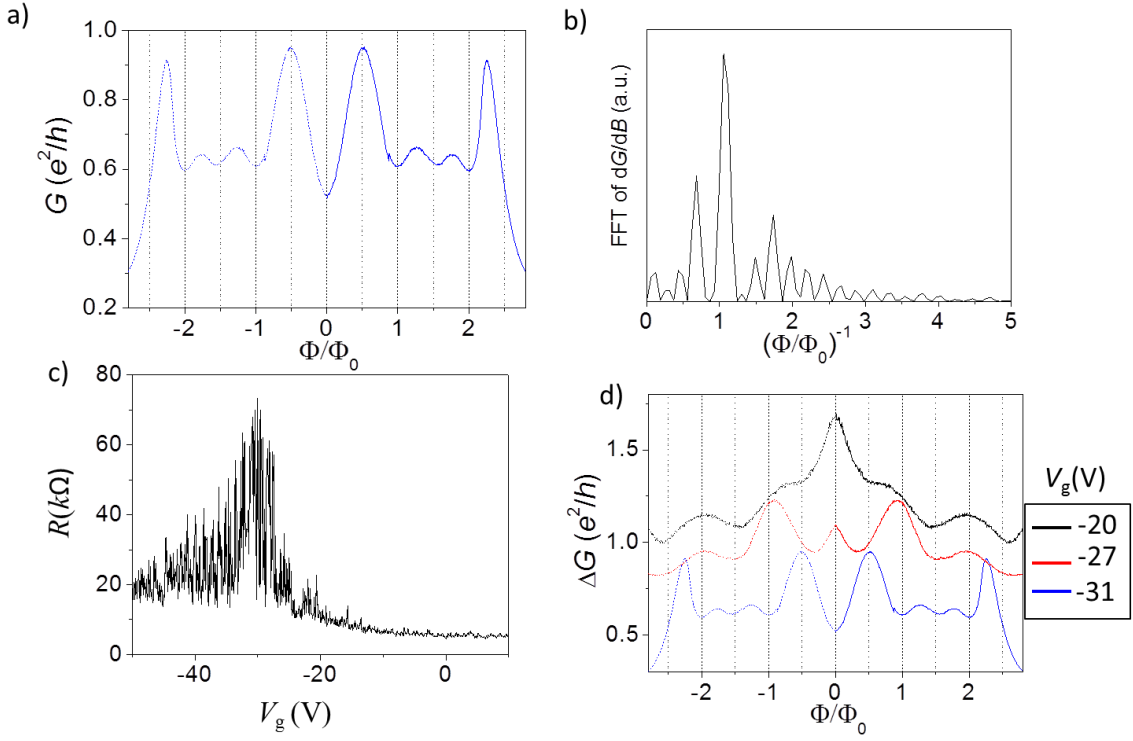

**Fig. S3| AB oscillation in device 2.** (a) Magneto-conductance at the Dirac point (Vg = -31V), showing a conductance minimum at $\Phi/\Phi_0 = 0$. (b) FFT of dG/dB from the data in (a). (c) Gate dependent four-point resistance, showing Dirac point. (d) Magneto-conductance measured for increasing gate voltages, showing alternating AB phase.

Supplementary Discussion C: Model Hamiltonian and Relevant Parameters

Here we present details of the model Hamiltonian used in the manuscript to create Fig. 4. Computing time limitations necessitate the use of a simplified TI model having parameters that are somewhat different from those in the experiment. Within this constraint, we have made sure that the salient parameters (such as the surface state subband gap and nanowire aspect ratio) are reasonably close to those of the experiment, and that the simulations do not depend strongly on exact parameter values (discussed in more detail below).

We consider an effective cubic-lattice model for the 3D Topological Insulator (TI) Hamiltonian with two bands (A, B) and two spin flavors (↑, ↓). The resultant four-orbital Dirac Hamiltonian for the tight-binding model can be written as[6]

$$H_D = \sum_{\vec{r}} \left\{ (M\Gamma^0 - \mu \mathbb{I}_4)\Psi_{\vec{r}}^\dagger \Psi_{\vec{r}} + \sum_\delta \frac{1}{2}(B\Gamma^0 + iA\Gamma^\delta)\Psi_{\vec{r}}^\dagger \Psi_{\vec{r}+\delta} \right\}, \quad (S1)$$

where $\Psi_{\vec{r}} = (c_{A,\uparrow,\vec{r}} c_{A,\downarrow,\vec{r}} c_{B,\uparrow,\vec{r}} c_{B,\downarrow,\vec{r}})^T$ is the four-component spinor involving the spin and orbital degrees of freedom over all cubic lattice sites, $\vec{r} = (i, j, k)a_0$, which denotes lattice location with integers $i, j, k$ and lattice constant $a_0 = 1$ Å. Only nearest neighbor hopping is considered, so that $\delta = \pm a_0 \hat{x}, \pm a_0 \hat{y}, \pm a_0 \hat{z}$. The 3D system has length, width, and thickness dimensions of $(N_x, N_y, N_z)a_0$. We set $(N_x, N_y, N_z) = (24, 8, 8)$ in our system so that computational time is reasonable while retaining an aspect ratio and a surface state sub band gap energy reasonably close to experimental values.

The Gamma matrices are definied as $\Gamma^\eta = \tau^x \otimes \sigma^\eta$ and $\Gamma^0 = \tau^z \otimes \mathbb{I}_2$. $\tau^\eta$ and $\sigma^\eta$ are the 2 x 2 Pauli matrices representing orbital and spin degrees of freedom, $\eta \in x, y, z$. $\mathbb{I}_N$ is the N x N identity matrix, and chemical potential $\mu$. The parameters $A$ and $B$ determine the band structure, and are set to $A = 1$ eV· Å and $B = 1$ eV· Å², so that the model is isotropic and preserves time reversal and particle-hole symmetry. The mass term $M = m - 3\frac{B}{a_0^2}$ yields a topologically nontrivial (trivial) phase when $m/B > 0$ ($m/B < 0$), and is set to $m = 1.5$ eV in the topologically nontrivial model. While the system parameters used do not quantitatively mimic the band structure of Bi$_2$Se$_3$, the model is necessary to pin the surface state modes more closely to the surface so that surface state hybridization does not play a role in the narrow and thin nanowire.

The phenomenological expression for the self-energy term of the metallic contacts, with uniform density of states profile across energy, is[6]

$$\Sigma^r(\omega) = -i\Gamma \mathbb{I}_{4N_yN_z}, \quad (S2)$$

where tunneling rate $\Gamma = 1$ results in transparent contacts with the nanowire. The contacts are connected to the left-most ($x = 1$) and right-most ($x = N_x$) faces of the nanowire, such that the size of each self-energy matrix is $4 \times N_y \times N_z$ (4 for two sublattice and two spin degrees of freedom).

The magnetic field effect can be introduced to the Hamiltonian in terms of a nonzero vector potential, $\vec{A}(\vec{r})$, which gives rise to an additional phase factor relative to the real-space Hamiltonian with no vector potential. This is added via a Peierls substitution[7],

$$\langle \vec{r} | H | \vec{r'} \rangle \Rightarrow \langle \vec{r} | H | \vec{r'} \rangle \times e^{i \frac{q}{\hbar} \int_{\vec{r}}^{\vec{r'}} \vec{A}(\vec{l}) \cdot d\vec{l}}, \tag{S3}$$

where $\vec{l}$ is the path from $\vec{r}$ to $\vec{r'}$. We focus on the effect of a homogeneous magnetic field lying in the longitudinal direction, $\vec{B} = \hat{x} B_x$, which in the Landau gauge yields a vector potential $\vec{A} = -\hat{y} B_x z$ that is only nonzero in the $\hat{y}$ direction. Lattice hopping in the width direction is thus the only component in the Hamiltonian that accrues an additional phase,

$$\langle \vec{r} | H | \vec{r} + \hat{y} a_0 \rangle \Rightarrow \langle \vec{r} | H | \vec{r} + \hat{y} a_0 \rangle \times e^{i 2\pi (z-1) \Phi_p / \Phi_0}, \tag{S4}$$

where $\Phi_p$ is the magnetic flux threaded through each cross-sectional plaquette of area $a_0^2$, and $\Phi_0 = h/q$ is the magnetic flux quantum. Total magnetic flux through the nanowire is $\Phi = N_y N_z \Phi_p$. Zeeman splitting is roughly an order of magnitude smaller than surface state sub band gap in the experimental regime on which we focus ($\Phi/\Phi_0 < 1$), and as such is neglected in the simulations.

This model thus provides a framework for studying the Aharanov-Bohm effect in TI nanowires by calculating the transmission (i.e. the differenctial conductance) from the left metallic contact to the right metallic contact, as a function of both chemical potential and longitudinal magnetic field. Transmission is a conventional observable calculated within the non-equilibrium Green Function (NEGF) formalism[8].